\documentclass[aps,prl,showpacs,superscriptaddress,twocolumn]{revtex4-1}
\usepackage[english]{babel} \usepackage{graphicx} \usepackage{stmaryrd}
\usepackage{amsmath, amssymb} \usepackage{mathrsfs} \usepackage{xspace}
\usepackage{bm} \usepackage{color} \definecolor{lightblue}{rgb}{0.13,
0.26, 0.99}

\usepackage[dvipdfmx, colorlinks=true, urlcolor=lightblue,
citecolor=lightblue, linkcolor=lightblue,
hyperfootnotes=false]{hyperref}

\newcommand{\Tr}{\operatorname{Tr}}

\begin{document}

\title{Symmetry Protection of Critical Phases and a Global Anomaly in
$1+1$ Dimensions}

\author{Shunsuke C. Furuya}
\affiliation{Condensed Matter Theory Laboratory, RIKEN, Wako, Saitama 351-0198, Japan}
\author{Masaki Oshikawa}
\affiliation{Institute for Solid
State Physics, University of Tokyo, Kashiwa 277-8581, Japan}
\date{\today}
\begin{abstract}
We derive a selection rule among the $(1+1)$-dimensional SU(2)
Wess-Zumino-Witten theories,
based on the global anomaly of the discrete
$\mathbb{Z}_2$ symmetry found by Gepner and Witten.
In the presence of both the SU(2) and $\mathbb{Z}_2$ symmetries,
a renormalization-group flow is possible between level-$k$ and level-$k'$
Wess-Zumino-Witten theories only if $k\equiv k' \mod{2}$.
This classifies the Lorentz-invariant, SU(2)-symmetric
critical behavior into two ``symmetry-protected'' categories
corresponding to even and odd levels,
restricting possible gapless critical behavior of
translation-invariant quantum spin chains.
\end{abstract}
\maketitle

\textit{Introduction.---}
The classification of quantum phases is a central problem in condensed-matter and statistical physics.
They are first classified into
gapped and gapless phases.
There has been a significant
progress in further classification of gapped phases,
which are relatively easy
to be handled theoretically.
In particular, symmetry protected
topological (SPT) phases~\cite{GuWen-SPT-PRB2009,Pollmann_2012}
have become an important concept in the
classification.
That is, even when two states have no long-range entanglement and are
indistinguishable in terms of any local observables, they could still
belong to distinct phases separated by a quantum phase transition, in
the presence of a certain symmetry.  In contrast, classification of
gapless quantum phases remains very much open.  Symmetries are naturally
expected to play an important role also in the classification of gapless
quantum phases.

The issue of classification of quantum phases is also deeply
related to that of quantum field theories.
Since a quantum field theory can be regarded as an effective description
of universal low-energy behaviors of quantum many-body systems,
we may expect that they are essentially the same problem.
An intriguing feature of quantum field theory is the \emph{anomaly}~\cite{A_anomaly,BJ_anomaly,Witten_SU2_anomaly,Kapustin_anomaly}.
A particularly interesting anomaly is the global anomaly, which emerges after promotion of a global symmetry to gauge symmetry~\cite{Kapustin_anomaly}.
As it is the case with many connections between quantum field theory
and condensed-matter physics, investigation of the consequences
of the anomaly in quantum field theory in a condensed-matter physics
context has been quite fruitful, including the discovery of
the ``Chern insulator''~\cite{Haldane1988}.
Nevertheless, the exact correspondence in many concrete cases is not yet understood.

An anomaly in a field theory may imply that the field theory
cannot be realized in a condensed-matter system or a lattice
model in the same dimensions.
A renowned case is the impossibility of realization of
a chiral fermion with (noninteracting) fermions on a lattice,
known as the Nielsen-Ninomiya theorem~\cite{NielsenNinomiya1,NielsenNinomiya2},
which is deeply related to
the chiral anomaly, a representative example of the quantum anomaly.
Nevertheless, such a field theory could be realized at the boundary (edge or surface) of a condensed-matter system
in higher dimensions. For example,
the chiral fermion in $1+1$ dimensions indeed appears as
the edge state of a two-dimensional quantum Hall system.
Ryu \textit{et al.}~\cite{Ryu_modular_invariance,Sule_orbifold}
generalized this observation to classification of gapped SPT phases:
the edge or surface state of an SPT phase exhibits an anomaly with
respect to the relevant
symmetry, which implies the ``ingappability'' of the edge state in
the presence of the symmetry~\cite{Hsieh_orientifold}.
Conversely, such an anomaly can be identified with an edge or surface
state, and thus with an SPT phase in higher dimensions.

These developments motivate us to question if there is a mechanism of
symmetry protection of the universality class of \emph{bulk} gapless
critical phases.
In this Letter, we argue that there is a protection of bulk gapless
critical phases by discrete symmetry.  This symmetry protection is
analogous to that of the well-known (gapped) SPT phases; here we
show that the concept can be generalized to bulk \emph{gapless critical}
phases.
We demonstrate this for the SU(2)-symmetric quantum
antiferromagnetic chains and their effective field theory,
SU(2) Wess-Zumino-Witten (WZW) theory as
an example.
The SU(2) WZW theory is characterized by a non-negative integer
$k$, which is called level.  Hereafter we denote the level-$k$ SU(2) WZW
theory as WZW$_k$.  
WZW$_k$'s with $k=1,2,\ldots$ are 
a complete classification
of the universality classes of critical points in $1+1$ dimensions
with the Lorentz and SU(2) symmetry only.
We can also identify 
WZW$_0$ with a gapped phase with a unique ground state.

Our main claim in the present Letter is as follows:
in the presence of the SU(2) and
a certain discrete
$\mathbb{Z}_2$ symmetry of the WZW theory, which corresponds to
the translation symmetry of the spin chain, 
a renormalization-group (RG) flow is possible between WZW$_k$ and
WZW$_{k'}$ only if $k\equiv k'\mod{2}$.  That is, the
gapless critical phases in $1+1$ dimension with the SU(2), the
$\mathbb{Z}_2$, and the Lorentz symmetries are classified into two
``symmetry-protected'' categories: one corresponds to even levels and
the other to odd levels.
In terms of spin chains, as long as 
the SU(2) spin rotation and the lattice translation symmetries
are unbroken (either explicitly or spontaneously),
a spin chain with $S\in\mathbb{Z}$ can only realize WZW$_k$
with an even $k$,
while one with $S\in\mathbb{Z}+1/2$ can only realize WZW$_k$ with an odd $k$.

Our argument is based on the global anomaly of a discrete symmetry
in the WZW theory originally found by
Gepner and Witten in 1986~\cite{Gepner_wzw},
providing a new link between the anomaly in quantum field theory and
condensed-matter physics.
As we will discuss 
later, the present result includes, as
special cases, the earlier semi-classical ($k\to\infty$)
analysis~\cite{AffleckHaldane}, and the Lieb-Schultz-Mattis
theorem~\cite{LSM,AffleckLieb,Xu_NLSM} applied to SU(2)-symmetric
one-dimensional systems.  However, the present result is much stronger
than the Lieb-Schultz-Mattis theorem, in restricting the possible
universality class of the gapless critical phase.
Furthermore, we shall discuss an experimental consequence of the
present result, in terms of Raman spectroscopy as an example.

\textit{Model.---}
The standard Heisenberg antiferromagnetic (HAFM)
chain is defined by the Hamiltonian
$\mathcal{H}_{\rm HAFM}=J_1\sum_j\bm{S}_j\cdot\bm{S}_{j+1}$
with $J_1>0$,
which possesses the SU(2) symmetry of the global spin rotation, the
lattice translation symmetry, and the lattice inversion symmetry.  We
can also consider various generalizations of this model by including
next-nearest-neighbor interaction~\cite{Pixley_Spin1}, biquadratic interaction~\cite{Kitazawa}, and so on~\cite{Michaud_dimer,Wang_3spin}.

In order to explore the possible quantum critical
behavior of quantum spin chains,
non-Abelian bosonization is useful.
In the non-Abelian bosonization of quantum antiferromagnetic chains,
first the spin-$S$ spin chain is represented in terms of fermions with
$2S$ ``colors''~\cite{AffleckHaldane}.
The resulting effective field theory is 
WZW$_k$
defined by the action
\begin{equation}
\mathcal S_{k} = -\frac 1{\lambda}\int_{S^2} dx_0dx_1 \Tr
  [(g^{-1}\partial_\mu g)(g^{-1} \partial_\mu g)] + k\Gamma_{\rm wz},
  \label{action}
\end{equation}
with the $\mathrm{SU(2)}$ matrix field  $g(x_0,x_1)$ with the spin indices,
the coupling constant $\lambda>0$,
and the Wess-Zumino term $k\Gamma_{\mathrm{WZ}}$.
We consider the space-(imaginary) time compactified as the
two-dimensional sphere $S^2$.
The Wess-Zumino term $\Gamma_{\mathrm{WZ}}$
is defined on the three-dimensional ball $B^3$,
extended from the spacetime manifold $S^2$.
It thus appears to depend on the extension of $g$ to $B^3$, which
is arbitrary.
However, it is a topological term unaffected by any infinitesimal variation of the extension.
Nevertheless, $\Gamma_{\rm WZ}$ can take values different by
integral multiples of $2\pi$, corresponding to topologically
inequivalent extensions.  
To define the theory consistently, the partition function should
be independent of the arbitrary extension to $B^3$, and thus
the level $k$ is quantized to be
$k\in\mathbb{Z}$~\cite{Witten_nonabelian_bosonization}.
WZW$_k$ is a conformal field theory (CFT)
with the SU(2) symmetry, for each level $k$.
Exact features of WZW$_k$ are known, thanks
to its infinite dimensional symmetry governed
by Kac-Moody algebra~\cite{Witten_nonabelian_bosonization,Gepner_wzw}. 
Each value of $k$ represents the different critical behaviors.

In the non-Abelian bosonization treatment of spin-$S$ chains~\cite{AffleckHaldane}, the level is naturally
given as $k=2S$.
Indeed, an integrable model called Takhtajan-Babujian model is
known for each $S$, and its exact solution shows that its low-energy
limit is described by WZW$_{2S}$.
However, generically the effective theory of the spin-$S$
chain contains various perturbations to WZW$_{2S}$.
The general principle is that, all the possible
perturbations allowed by the symmetries should be present, unless
parameters in the Hamiltonian are fine-tuned.
In fact, the Takhtajan-Babujian model corresponds
to the special multicritical point where the parameters are fine-tuned
so that all the relevant perturbations vanish.
More generic models usually have
relevant perturbations which
drive the system away from the original WZW$_{2S}$ fixed point,
under RG.

To
discuss the RG flow, we need to identify symmetries of the
system and their representation in the field theory.  In this Letter, we
limit ourselves to models with the global SU(2) symmetry of spin
rotation.  Furthermore, we consider the models which are invariant
under the translation by one site, $T_1:\bm{S}_j\to\bm{S}_{j+1}$.
The lattice translation symmetry $T_1$ is represented by the
$\mathbb{Z}_2$ symmetry under $g\to-g$, 
in the WZW theory~\cite{AffleckHaldane}.

\textit{Modular invariance.---}
To see the consistency of a CFT, it is convenient to
consider the system on a torus.
The torus can be defined in terms of
complex coordinates $z$ and $\bar{z}$ with the identifications $z\sim z+2\pi$ and $z\sim z+2\pi\tau$ with the modulus $\tau\in\mathbb{C}$.
The conformal invariance enhanced by the SU(2) symmetry (Kac-Moody
algebra) dictates that the partition function on the torus is
generally given as
\begin{equation}
 Z(\tau,\bar\tau) = \sum_{j,j'=0}^{k/2}
  \chi_j(\tau)X_{j,j'}\bar\chi_{j'}(\bar\tau), \label{Z_chi}
\end{equation}
where  $\chi_j(\tau)$ and $\bar\chi_{j'}(\bar\tau)$ are Kac-Moody characters as functions of
the modulus $\tau$, corresponding to holomorphic and anti-holomorphic
parts~\cite{DiFrancesco_CFT}. 
The characters are labeled by the ``spin'' $j=0,1/2,\ldots,k/2$.
The coefficients $X_{j,j'}$ count the number of the primary field with the spin $(j,j')$.
Thus $X_{j,j'}$ must be non-negative integers in a consistent CFT.

The same torus can be represented by different modular parameters,
which are related by modular transformations generated by
$\mathcal{T}:\quad\tau\to\tau+1$ and
$\mathcal{S}:\quad\tau\to-1/\tau$.
Here $\mathcal{T}$ twists the spatial boundary condition and
$\mathcal{S}$ exchanges the space and imaginary-time directions~\cite{suppl}.
Since these keep the underlying torus unchanged,
a physically sensible partition function should be invariant
under them.
This requirement, which is called modular invariance, in fact
leads to quite a powerful constraint on a possible consistent CFT.
In the present context, $X_{j,j'}$ are strongly constrained by
the modular invariance so as to be a non-negative integer~\cite{DiFrancesco_CFT, suppl}.

The standard partition function of WZW$_k$ on the torus with the
periodic boundary conditions is given by
\begin{align}
Z_{\mathrm{SU(2)}} (\tau,\bar\tau)= 
\Tr{\left( e^{-2\pi \operatorname{Im}{\tau} \mathcal H}e^{i2\pi
\operatorname{Re}{\tau} \mathcal P}\right)}, \label{tZ_SU2}
\end{align}
where $\mathcal{H}$ and $\mathcal{P}$ are the
total energy and momentum, respectively.
This indeed turns out to be modular invariant partition function
with $X_{j,j'}=\delta_{j,j'}$.

However, this is not the only possible modular invariant partition
function.
Since WZW$_k$ also possesses the discrete $\mathbb{Z}_2$ symmetry
$g\to-g$, we also consider projecting the Hilbert space
onto the subspace which is symmetric under $g \to -g$.
The resulting partition function reads
\begin{align}
Z^{\mathrm{proj}}_+ (\tau,\bar\tau)= 
\Tr{\left( P_+ e^{-2\pi \operatorname{Im}{\tau} \mathcal H}e^{i2\pi
\operatorname{Re}{\tau} \mathcal P}\right)},
\label{tZ_+}
\end{align}
where $P_+$ is the projection operator onto the
subspace which is even under $g\to-g$.
In the path-integral formalism, the insertion of $P_+$ is equivalent
to averaging over the periodic and the antiperiodic boundary conditions
on $g$ in the imaginary time direction.
As a consequence, $Z^{\mathrm{proj}}_+$ is \emph{not} modular invariant.
We can construct a modular invariant partition function based
on $Z^{\mathrm{proj}}_+$, by ``gauging'' the $\mathbb{Z}_2$ symmetry
also in the spatial direction.
This procedure is known as orbifold construction, and the resulting
partition function of the $\mathbb{Z}_2$ orbifold of 
WZW$_k$ reads
\begin{equation}
 Z_+(\tau,\bar\tau)
= (1+\mathcal S + \mathcal T\mathcal S)Z^{\rm
  proj}_+(\tau,\bar\tau) - Z_{\mathrm{SU(2)}}(\tau,\bar\tau).
  \label{Z_+_ST}
\end{equation}
Despite the construction to make it modular invariant,
in fact, $Z_+(\tau,\bar\tau)$
is modular invariant only if $k$ is even; it is modular noninvariant if
$k$ is odd~\cite{Gepner_wzw,suppl}.
This is an example of a global anomaly in
quantum field theory.
While the presence of this anomaly can be interpreted by
a construction of SO(3) WZW theory~\cite{Gepner_wzw} as
SU(2) modulo $g \sim -g$ is SO(3),
the physical implications of the anomaly has not been elucidated.

\textit{Consequences of the global anomaly.---}
Now we shall argue that there are indeed very profound consequences.
First, we consider a RG flow from WZW$_k$ to WZW$_{k'}$
induced by perturbations allowed within the SU(2) and
the $\mathbb{Z}_2$ ($g\to-g$) symmetries, for an even $k$.
While both fixed points have the SU(2) Kac-Moody and $\mathbb{Z}_2$
symmetries for any $k$ and $k'$, the level $k'$ of the infrared
fixed point is not arbitrary.
Since $k$ is assumed to be even, for the ultraviolet fixed point
WZW$_k$, the $\mathbb{Z}_2$ orbifold
is a consistent conformal field theory with a modular
invariant partition function.
Because the orbifold construction is given by summation over
partition functions with various boundary conditions,
it should not affect the RG flow in the bulk.
In other words, once the projection onto the symmetric subspace
is done consistently, the RG flow can be followed under the projection.
Thus the RG flow between WZW theories under the $\mathbb{Z}_2$ symmetry
implies a corresponding RG flow between their $\mathbb{Z}_2$ orbifolds.
This means that, the infrared fixed point WZW$_{k'}$ should have a
consistent $\mathbb{Z}_2$ orbifold and thus $k'$ must be even.

Next we consider the RG flow from WZW$_k$ to WZW$_{k'}$, when
$k$ is odd.
Now the ultraviolet fixed point has the global anomaly and
the $\mathbb{Z}_2$ orbifold is ill defined.
There is no \emph{a priori} reason that the infrared fixed point has
the same anomaly, since symmetries can, generally speaking,
emerge in the infrared limit.
Nevertheless, below we shall argue that the global anomaly in the
ultraviolet fixed point is ``inherited'' by the infrared limit.

Naively, since the theory has the discrete $\mathbb{Z}_2$
symmetry $g\to-g$,
we expect that we can consider projection of the entire
Hilbert space onto the subspace which is symmetric
under $g\to-g$.
However the anomaly for odd $k$ precisely means that
there is no consistent CFT defined
within the symmetric (or antisymmetric) subspace.
The odd-$k$ WZW$_k$ is inconsistent unless both symmetric
and antisymmetric sectors are included.
This observation can be related to the ``gappability'' of the theory.
In general, a CFT has relevant operators
in its spectrum. Once the theory is perturbed by a relevant
operator, generically the theory would become massive;
the excited states would be separated from the ground state
by a nonvanishing mass gap.
Usually the ground state in such a system is unique.
If this is the case, 
because of the $\mathbb{Z}_2$ symmetry of the theory,
the unique ground state is either symmetric or antisymmetric
with respect to the $\mathbb{Z}_2$ symmetry.
Then, in order to describe the low-energy physics,
we can consider a projection onto
symmetric or antisymmetric sector of the Hilbert space.
However, for odd $k$, the global anomaly of WZW$_k$
means that such a projection does not yield a consistent
quantum field theory, 
Therefore, in order to open a mass gap, the global anomaly
for odd $k$ requires that the ground states below the gap
exist in both the symmetric and antisymmetric sectors
and are doubly degenerate.
This signals the spontaneous breaking of 
the $g\to-g$ symmetry~\footnote{
In general, the double degeneracy of the ground states
in the symmetric
and antisymmetric sectors with respect to the $g\to-g$
does not necessarily mean the spontaneous symmetry breaking:
it could correspond to a topological degeneracy 
two ground states in the presence of periodic boundary conditions
are indistinguishable by any local observable.
However, in one spatial dimension, which is the focus
of the present Letter, such a topological order 
with long-range entanglement 
is absent~\protect\cite{Chen_classification_1D_PRB2011}
and thus the ground-state degeneracy between the two sectors
does imply a spontaneous symmetry
breaking.}.
The above statement corresponds to a field-theory version of
the Lieb-Schultz-Mattis theorem~\cite{LSM,AffleckLieb,Xu_NLSM},
as it will become clearer in the correspondence with spin chains
discussed below.

Now suppose there is an RG flow from WZW$_k$ with an odd $k$ to WZW$_{k'}$
with an even $k'$,
we can further perturb the infrared fixed point
to obtain a massive (gapped) field theory with a unique
ground state (corresponding to WZW$_0$).
This contradicts with the ingappability of WZW$_k$
discussed above. Thus $k'$ must also be odd.
Combining the results for even and odd $k$, we obtain the following statement.

\emph{When there is an RG flow from WZW$_k$ to WZW$_{k'}$, if the SU(2) symmetry
and the $\mathbb{Z}_2$ symmetry $g\to-g$ are respected,
$k'<k$ and $k'\equiv k\mod{2}$.}

In addition, it has been known that $k' < k$
thanks to Zamolodchikov's $c$-theorem~\cite{Zamolodchikov_c}, which dictates 
that the central charge of the infrared fixed point WZW$_{k'}$ 
should be smaller than that of the ultraviolet one WZW$_k$.

The implications of the above field-theory constraint
on spin chains is as follows.
As discussed earlier, non-Abelian bosonization of
a spin-$S$ HAFM chain yields WZW$_k$
with $k=2S$, with perturbations allowed by the symmetries.
Thus we find the following:

\emph{
The critical behavior of a general spin-$S$ HAFM chain
is described by WZW$_k$ with
$k\equiv 2S\mod{2}$, as long as the Hamiltonian
possesses the SU(2) spin rotation symmetry and
the lattice translation symmetry.}

In other words, critical phenomena in 1+1 dimensions
with SU(2) and $\mathbb{Z}_2$ symmetry of $g\to-g$
are grouped into two symmetry-protected classes:
one consists of WZW$_k$ with even $k$
and the other with odd $k$.
In the presence of the  SU(2) spin rotation symmetry
and the lattice translation symmetry,
general HAFM chains with integer spin $S$
can only realize the former,
while those with half-odd-integer spin $S$
can only realize the latter.
The present argument can also be generalized to the case of the
site-centered inversion symmetry~\cite{suppl,Pradisi_unori1,Pradisi_unori2}.

Affleck and Haldane~\cite{AffleckHaldane} argued, based on the large-$k$
semiclassical analysis, WZW$_k$  with the leading perturbation
allowed under the $g \to -g$ symmetry,
$(\operatorname{tr}g)^2$, can be mapped
to the O(3) nonlinear sigma model at the topological angle $\theta=\pi k$.
Given that it is equivalent to WZW$_0$ or WZW$_1$ when $k$ is even or odd
respectively, their result is a special case of the present one.
On the other hand, the present result generalizes significantly
that of Ref.~\cite{AffleckHaldane} in restricting RG flows among
$k>1$ fixed points induced by generic perturbations.

\textit{Examples.---}
Let us now discuss several concrete cases, in light of the present result.

Ziman and Schulz studied translation invariant $S=3/2$
antiferromagnetic chains numerically and confirmed that,
while the Takhtajan-Babujian model is described by WZW$_3$,
away from the special Takhtajan-Babujian point
the system is described by
WZW$_1$~\cite{ZimanSchulz}. Namely, there is an RG flow
from WZW$_3$ to WZW$_1$, in accordance
with the present result.
The critical behavior corresponding to WZW$_2$
is not found in the model, although it is allowed by
the $c$-theorem.

Another interesting example is
the translation invariant spin-$S$ model~\cite{Michaud_dimer}
\begin{equation}
 \mathcal{H}_{\mbox{\scriptsize $J_1$-$J_3$}}=
\sum_j[J_1\bm{S}_j\cdot\bm{S}_{j+1}+J_3\{(\bm{S}_{j-1}\cdot\bm{S}_j)(\bm{S}_j\cdot\bm{S}_{j+1})+\mathrm{H.c.}\}].
  \label{J_1-J_3}
\end{equation}
This model has the WZW$_{2S}$ quantum
critical point $J_{3c}>0$~\cite{Michaud_wzw}, even though
the model is not at the integrable Takhtajan-Babujian point.
Again this is consistent with our selection rule.

An $S=1$ spin chain of Ref.~\cite{Pixley_Spin1} is a highly nontrivial example of our theory. 
It possesses a WZW$_4$ multicritical point whose level is higher than $2S$ and still consistent with our selection rule.

We stress that our selection rule is protected by the
one-site translation symmetry.
In fact, it can be removed by breaking the
translation symmetry explicitly.
For example, an extended model~\cite{Wang_3spin}
\begin{equation}
 \mathcal H_{\mbox{\scriptsize $J_1$-$J_3$-$\delta$}}=
\mathcal H_{\mbox{\scriptsize $J_1$-$J_3$}}
-J_1\delta\sum_j(-1)^j\bm S_j \cdot \bm S_{j+1} \label{J_1-J_3-delta}
\end{equation}
breaks the one-site translation symmetry explicitly, when $\delta \neq 0$.
When $S=1$, the model~\eqref{J_1-J_3-delta}
exhibits a critical line of $c=1$ connected to the
multicritical point with $c=3/2$.  This means that WZW$_2$
can flow to WZW$_1$ in the absence of the 
translation symmetry.  A similar RG flow from the level $2$ to the
level $1$ is found in the $S=1$ bilinear-biquadratic chain with the bond
alternation~\cite{Kitazawa}.
The bond alternation breaks the lattice translation and 
the site-centered inversion symmetries, but keeps the time reversal
and the bond-centered inversion symmetries.
This is consistent with our analysis that either the lattice
translation or the site-centered inversion symmetry protects
the two categories, while the time reversal nor the site-centered
inversion symmetry does not~\cite{suppl}.

\textit{Observable consequences.---}
The Raman spectroscopy is important in studying one-dimensional spin systems~\cite{Simutis_Raman_ladder},
probing dynamical correlation of
the Raman operator $R \sim \sum_j\bm{S}_j\cdot\bm{S}_{j+1}$ in
spin chains~\cite{Michaud_Raman,Sato_Raman}.
Dimensional analysis leads to the Raman spectrum
$I(\omega)\propto(\omega/J)^{2x-2}$
in the low energy limit, $\hbar\omega\ll k_BT\ll J$,
where $\omega$ is the Raman frequency shift, $J$ is the (typical)
spin-spin interaction energy scale, and $x$ is the scaling
dimension of $R$.
Near the WZW$_k$ quantum
critical point of a translation-invariant spin chain, we can derive
$R\propto\int dx\;(\operatorname{tr}g)^2$, which implies
$x=2/(2+k)$.
If the one-site translation symmetry is broken, 
$R\propto\int dx\;\operatorname{tr}g$ and 
$x=3/\left(4(2+k)\right)$.
Thus the Raman spectrum $I(\omega)$ provides a rather direct
probe of the level $k$ of the effective WZW$_k$ field theory
of the spin chain, and would also be useful in studying
crossover among WZW with different levels as discussed
in this Letter.

\textit{Conclusions.---}
In this Letter, we proposed the concept of symmetry protection of critical
phases. As an example, we demonstrated that the SU(2) WZW theories
are classified into two categories of even and odd
levels, in the presence of the discrete $\mathbb{Z}_2$ symmetry
of WZW theory which corresponds to the one-site translation symmetry
of spin chains.
The present result provides a new direction for the classification
of quantum phases, as well as a novel link between the anomaly in
field theory and condensed-matter physics.
It would be interesting to find more examples with different symmetries
or in higher dimensions.

\textit{Acknowledgments.---}
We are grateful to Soichiro Mohri, Andriy Nevidomskyy, and Kantaro Ohmori for
useful discussions.
S.C.F. was supported by the Swiss SNF under Division II.
The present work is supported in part by JSPS KAKENHI Grants
No. 16J04731 (S.C.F.) and No. 25400392 and No. 16K05469 (M.O.), and 
JSPS Strategic International Networks Program No. R2604 ``TopoNet''.
M. O. was also supported by US National Science Foundation grants PHY-1066293 and PHY-1125915, respectively, through Aspen Center for Physics and Kavli Institute for Theoretical Physics, UC Santa Barbara, where parts of this work were performed.

%


\begin{thebibliography}{34}%
\makeatletter
\providecommand \@ifxundefined [1]{%
 \@ifx{#1\undefined}
}%
\providecommand \@ifnum [1]{%
 \ifnum #1\expandafter \@firstoftwo
 \else \expandafter \@secondoftwo
 \fi
}%
\providecommand \@ifx [1]{%
 \ifx #1\expandafter \@firstoftwo
 \else \expandafter \@secondoftwo
 \fi
}%
\providecommand \natexlab [1]{#1}%
\providecommand \enquote  [1]{``#1''}%
\providecommand \bibnamefont  [1]{#1}%
\providecommand \bibfnamefont [1]{#1}%
\providecommand \citenamefont [1]{#1}%
\providecommand \href@noop [0]{\@secondoftwo}%
\providecommand \href [0]{\begingroup \@sanitize@url \@href}%
\providecommand \@href[1]{\@@startlink{#1}\@@href}%
\providecommand \@@href[1]{\endgroup#1\@@endlink}%
\providecommand \@sanitize@url [0]{\catcode `\\12\catcode `\$12\catcode
  `\&12\catcode `\#12\catcode `\^12\catcode `\_12\catcode `\%12\relax}%
\providecommand \@@startlink[1]{}%
\providecommand \@@endlink[0]{}%
\providecommand \url  [0]{\begingroup\@sanitize@url \@url }%
\providecommand \@url [1]{\endgroup\@href {#1}{\urlprefix }}%
\providecommand \urlprefix  [0]{URL }%
\providecommand \Eprint [0]{\href }%
\providecommand \doibase [0]{http://dx.doi.org/}%
\providecommand \selectlanguage [0]{\@gobble}%
\providecommand \bibinfo  [0]{\@secondoftwo}%
\providecommand \bibfield  [0]{\@secondoftwo}%
\providecommand \translation [1]{[#1]}%
\providecommand \BibitemOpen [0]{}%
\providecommand \bibitemStop [0]{}%
\providecommand \bibitemNoStop [0]{.\EOS\space}%
\providecommand \EOS [0]{\spacefactor3000\relax}%
\providecommand \BibitemShut  [1]{\csname bibitem#1\endcsname}%
\let\auto@bib@innerbib\@empty
\bibitem [{\citenamefont {Gu}\ and\ \citenamefont
  {Wen}(2009)}]{GuWen-SPT-PRB2009}%
  \BibitemOpen
  \bibfield  {author} {\bibinfo {author} {\bibfnamefont {Z.-C.}\ \bibnamefont
  {Gu}}\ and\ \bibinfo {author} {\bibfnamefont {X.-G.}\ \bibnamefont {Wen}},\
  }\href {\doibase 10.1103/PhysRevB.80.155131} {\bibfield  {journal} {\bibinfo
  {journal} {Phys. Rev. B}\ }\textbf {\bibinfo {volume} {80}},\ \bibinfo
  {pages} {155131} (\bibinfo {year} {2009})}\BibitemShut {NoStop}%
\bibitem [{\citenamefont {Pollmann}\ \emph {et~al.}(2012)\citenamefont
  {Pollmann}, \citenamefont {Berg}, \citenamefont {Turner},\ and\ \citenamefont
  {Oshikawa}}]{Pollmann_2012}%
  \BibitemOpen
  \bibfield  {author} {\bibinfo {author} {\bibfnamefont {F.}~\bibnamefont
  {Pollmann}}, \bibinfo {author} {\bibfnamefont {E.}~\bibnamefont {Berg}},
  \bibinfo {author} {\bibfnamefont {A.~M.}\ \bibnamefont {Turner}}, \ and\
  \bibinfo {author} {\bibfnamefont {M.}~\bibnamefont {Oshikawa}},\ }\href
  {\doibase 10.1103/PhysRevB.85.075125} {\bibfield  {journal} {\bibinfo
  {journal} {Phys. Rev. B}\ }\textbf {\bibinfo {volume} {85}},\ \bibinfo
  {pages} {075125} (\bibinfo {year} {2012})}\BibitemShut {NoStop}%
\bibitem [{\citenamefont {Adler}(1969)}]{A_anomaly}%
  \BibitemOpen
  \bibfield  {author} {\bibinfo {author} {\bibfnamefont {S.~L.}\ \bibnamefont
  {Adler}},\ }\href {\doibase 10.1103/PhysRev.177.2426} {\bibfield  {journal}
  {\bibinfo  {journal} {Phys. Rev.}\ }\textbf {\bibinfo {volume} {177}},\
  \bibinfo {pages} {2426} (\bibinfo {year} {1969})}\BibitemShut {NoStop}%
\bibitem [{\citenamefont {Bell}\ and\ \citenamefont
  {Jackiw}(1969)}]{BJ_anomaly}%
  \BibitemOpen
  \bibfield  {author} {\bibinfo {author} {\bibfnamefont {J.}~\bibnamefont
  {Bell}}\ and\ \bibinfo {author} {\bibfnamefont {R.}~\bibnamefont {Jackiw}},\
  }\href {\doibase 10.1007/BF02823296} {\bibfield  {journal} {\bibinfo
  {journal} {Nuovo Cimento A}\ }\textbf {\bibinfo {volume} {60}},\ \bibinfo
  {pages} {47} (\bibinfo {year} {1969})}\BibitemShut {NoStop}%
\bibitem [{\citenamefont {Witten}(1982)}]{Witten_SU2_anomaly}%
  \BibitemOpen
  \bibfield  {author} {\bibinfo {author} {\bibfnamefont {E.}~\bibnamefont
  {Witten}},\ }\href {\doibase http://dx.doi.org/10.1016/0370-2693(82)90728-6}
  {\bibfield  {journal} {\bibinfo  {journal} {Phys. Lett.}\ }\textbf {\bibinfo
  {volume} {B117}},\ \bibinfo {pages} {324 } (\bibinfo {year}
  {1982})}\BibitemShut {NoStop}%
\bibitem [{\citenamefont {Kapustin}\ and\ \citenamefont
  {Thorngren}(2014)}]{Kapustin_anomaly}%
  \BibitemOpen
  \bibfield  {author} {\bibinfo {author} {\bibfnamefont {A.}~\bibnamefont
  {Kapustin}}\ and\ \bibinfo {author} {\bibfnamefont {R.}~\bibnamefont
  {Thorngren}},\ }\href {\doibase 10.1103/PhysRevLett.112.231602} {\bibfield
  {journal} {\bibinfo  {journal} {Phys. Rev. Lett.}\ }\textbf {\bibinfo
  {volume} {112}},\ \bibinfo {pages} {231602} (\bibinfo {year}
  {2014})}\BibitemShut {NoStop}%
\bibitem [{\citenamefont {Haldane}(1988)}]{Haldane1988}%
  \BibitemOpen
  \bibfield  {author} {\bibinfo {author} {\bibfnamefont {F.~D.~M.}\
  \bibnamefont {Haldane}},\ }\href {\doibase 10.1103/PhysRevLett.61.2015}
  {\bibfield  {journal} {\bibinfo  {journal} {Phys. Rev. Lett.}\ }\textbf
  {\bibinfo {volume} {61}},\ \bibinfo {pages} {2015} (\bibinfo {year}
  {1988})}\BibitemShut {NoStop}%
\bibitem [{\citenamefont {Nielsen}\ and\ \citenamefont
  {Ninomiya}(1981{\natexlab{a}})}]{NielsenNinomiya1}%
  \BibitemOpen
  \bibfield  {author} {\bibinfo {author} {\bibfnamefont {H.}~\bibnamefont
  {Nielsen}}\ and\ \bibinfo {author} {\bibfnamefont {M.}~\bibnamefont
  {Ninomiya}},\ }\href {\doibase
  http://dx.doi.org/10.1016/0550-3213(81)90524-1} {\bibfield  {journal}
  {\bibinfo  {journal} {Nucl. Phys.}\ }\textbf {\bibinfo {volume} {B193}},\
  \bibinfo {pages} {173 } (\bibinfo {year} {1981}{\natexlab{a}})}\BibitemShut
  {NoStop}%
\bibitem [{\citenamefont {Nielsen}\ and\ \citenamefont
  {Ninomiya}(1981{\natexlab{b}})}]{NielsenNinomiya2}%
  \BibitemOpen
  \bibfield  {author} {\bibinfo {author} {\bibfnamefont {H.}~\bibnamefont
  {Nielsen}}\ and\ \bibinfo {author} {\bibfnamefont {M.}~\bibnamefont
  {Ninomiya}},\ }\href {\doibase
  http://dx.doi.org/10.1016/0550-3213(81)90361-8} {\bibfield  {journal}
  {\bibinfo  {journal} {Nucl. Phys.}\ }\textbf {\bibinfo {volume} {B185}},\
  \bibinfo {pages} {20 } (\bibinfo {year} {1981}{\natexlab{b}})}\BibitemShut
  {NoStop}%
\bibitem [{\citenamefont {Ryu}\ and\ \citenamefont
  {Zhang}(2012)}]{Ryu_modular_invariance}%
  \BibitemOpen
  \bibfield  {author} {\bibinfo {author} {\bibfnamefont {S.}~\bibnamefont
  {Ryu}}\ and\ \bibinfo {author} {\bibfnamefont {S.-C.}\ \bibnamefont
  {Zhang}},\ }\href {\doibase 10.1103/PhysRevB.85.245132} {\bibfield  {journal}
  {\bibinfo  {journal} {Phys. Rev. B}\ }\textbf {\bibinfo {volume} {85}},\
  \bibinfo {pages} {245132} (\bibinfo {year} {2012})}\BibitemShut {NoStop}%
\bibitem [{\citenamefont {Sule}\ \emph {et~al.}(2013)\citenamefont {Sule},
  \citenamefont {Chen},\ and\ \citenamefont {Ryu}}]{Sule_orbifold}%
  \BibitemOpen
  \bibfield  {author} {\bibinfo {author} {\bibfnamefont {O.~M.}\ \bibnamefont
  {Sule}}, \bibinfo {author} {\bibfnamefont {X.}~\bibnamefont {Chen}}, \ and\
  \bibinfo {author} {\bibfnamefont {S.}~\bibnamefont {Ryu}},\ }\href {\doibase
  10.1103/PhysRevB.88.075125} {\bibfield  {journal} {\bibinfo  {journal} {Phys.
  Rev. B}\ }\textbf {\bibinfo {volume} {88}},\ \bibinfo {pages} {075125}
  (\bibinfo {year} {2013})}\BibitemShut {NoStop}%
\bibitem [{\citenamefont {Hsieh}\ \emph {et~al.}(2014)\citenamefont {Hsieh},
  \citenamefont {Sule}, \citenamefont {Cho}, \citenamefont {Ryu},\ and\
  \citenamefont {Leigh}}]{Hsieh_orientifold}%
  \BibitemOpen
  \bibfield  {author} {\bibinfo {author} {\bibfnamefont {C.-T.}\ \bibnamefont
  {Hsieh}}, \bibinfo {author} {\bibfnamefont {O.~M.}\ \bibnamefont {Sule}},
  \bibinfo {author} {\bibfnamefont {G.~Y.}\ \bibnamefont {Cho}}, \bibinfo
  {author} {\bibfnamefont {S.}~\bibnamefont {Ryu}}, \ and\ \bibinfo {author}
  {\bibfnamefont {R.~G.}\ \bibnamefont {Leigh}},\ }\href {\doibase
  10.1103/PhysRevB.90.165134} {\bibfield  {journal} {\bibinfo  {journal} {Phys.
  Rev. B}\ }\textbf {\bibinfo {volume} {90}},\ \bibinfo {pages} {165134}
  (\bibinfo {year} {2014})}\BibitemShut {NoStop}%
\bibitem [{\citenamefont {Gepner}\ and\ \citenamefont
  {Witten}(1986)}]{Gepner_wzw}%
  \BibitemOpen
  \bibfield  {author} {\bibinfo {author} {\bibfnamefont {D.}~\bibnamefont
  {Gepner}}\ and\ \bibinfo {author} {\bibfnamefont {E.}~\bibnamefont
  {Witten}},\ }\href {\doibase http://dx.doi.org/10.1016/0550-3213(86)90051-9}
  {\bibfield  {journal} {\bibinfo  {journal} {Nucl. Phys.}\ }\textbf {\bibinfo
  {volume} {B278}},\ \bibinfo {pages} {493 } (\bibinfo {year}
  {1986})}\BibitemShut {NoStop}%
\bibitem [{\citenamefont {Affleck}\ and\ \citenamefont
  {Haldane}(1987)}]{AffleckHaldane}%
  \BibitemOpen
  \bibfield  {author} {\bibinfo {author} {\bibfnamefont {I.}~\bibnamefont
  {Affleck}}\ and\ \bibinfo {author} {\bibfnamefont {F.~D.~M.}\ \bibnamefont
  {Haldane}},\ }\href {\doibase 10.1103/PhysRevB.36.5291} {\bibfield  {journal}
  {\bibinfo  {journal} {Phys. Rev. B}\ }\textbf {\bibinfo {volume} {36}},\
  \bibinfo {pages} {5291} (\bibinfo {year} {1987})}\BibitemShut {NoStop}%
\bibitem [{\citenamefont {Lieb}\ \emph {et~al.}(1961)\citenamefont {Lieb},
  \citenamefont {Schultz},\ and\ \citenamefont {Mattis}}]{LSM}%
  \BibitemOpen
  \bibfield  {author} {\bibinfo {author} {\bibfnamefont {E.}~\bibnamefont
  {Lieb}}, \bibinfo {author} {\bibfnamefont {T.}~\bibnamefont {Schultz}}, \
  and\ \bibinfo {author} {\bibfnamefont {D.}~\bibnamefont {Mattis}},\ }\href
  {http://www.sciencedirect.com/science/article/pii/0003491661901154}
  {\bibfield  {journal} {\bibinfo  {journal} {Ann. Phys. (N.Y.)}\ }\textbf
  {\bibinfo {volume} {16}},\ \bibinfo {pages} {407 } (\bibinfo {year}
  {1961})}\BibitemShut {NoStop}%
\bibitem [{\citenamefont {Affleck}\ and\ \citenamefont
  {Lieb}(1986)}]{AffleckLieb}%
  \BibitemOpen
  \bibfield  {author} {\bibinfo {author} {\bibfnamefont {I.}~\bibnamefont
  {Affleck}}\ and\ \bibinfo {author} {\bibfnamefont {E.}~\bibnamefont {Lieb}},\
  }\href@noop {} {\bibfield  {journal} {\bibinfo  {journal} {Lett. Math.
  Phys.}\ }\textbf {\bibinfo {volume} {12}},\ \bibinfo {pages} {57} (\bibinfo
  {year} {1986})}\BibitemShut {NoStop}%
\bibitem [{\citenamefont {Xu}\ and\ \citenamefont {Ludwig}(2013)}]{Xu_NLSM}%
  \BibitemOpen
  \bibfield  {author} {\bibinfo {author} {\bibfnamefont {C.}~\bibnamefont
  {Xu}}\ and\ \bibinfo {author} {\bibfnamefont {A.~W.~W.}\ \bibnamefont
  {Ludwig}},\ }\href {\doibase 10.1103/PhysRevLett.110.200405} {\bibfield
  {journal} {\bibinfo  {journal} {Phys. Rev. Lett.}\ }\textbf {\bibinfo
  {volume} {110}},\ \bibinfo {pages} {200405} (\bibinfo {year}
  {2013})}\BibitemShut {NoStop}%
\bibitem [{\citenamefont {Pixley}\ \emph {et~al.}(2014)\citenamefont {Pixley},
  \citenamefont {Shashi},\ and\ \citenamefont {Nevidomskyy}}]{Pixley_Spin1}%
  \BibitemOpen
  \bibfield  {author} {\bibinfo {author} {\bibfnamefont {J.~H.}\ \bibnamefont
  {Pixley}}, \bibinfo {author} {\bibfnamefont {A.}~\bibnamefont {Shashi}}, \
  and\ \bibinfo {author} {\bibfnamefont {A.~H.}\ \bibnamefont {Nevidomskyy}},\
  }\href {\doibase 10.1103/PhysRevB.90.214426} {\bibfield  {journal} {\bibinfo
  {journal} {Phys. Rev. B}\ }\textbf {\bibinfo {volume} {90}},\ \bibinfo
  {pages} {214426} (\bibinfo {year} {2014})}\BibitemShut {NoStop}%
\bibitem [{\citenamefont {Kitazawa}\ and\ \citenamefont
  {Nomura}(1999)}]{Kitazawa}%
  \BibitemOpen
  \bibfield  {author} {\bibinfo {author} {\bibfnamefont {A.}~\bibnamefont
  {Kitazawa}}\ and\ \bibinfo {author} {\bibfnamefont {K.}~\bibnamefont
  {Nomura}},\ }\href {\doibase 10.1103/PhysRevB.59.11358} {\bibfield  {journal}
  {\bibinfo  {journal} {Phys. Rev. B}\ }\textbf {\bibinfo {volume} {59}},\
  \bibinfo {pages} {11358} (\bibinfo {year} {1999})}\BibitemShut {NoStop}%
\bibitem [{\citenamefont {Michaud}\ \emph {et~al.}(2012)\citenamefont
  {Michaud}, \citenamefont {Vernay}, \citenamefont {Manmana},\ and\
  \citenamefont {Mila}}]{Michaud_dimer}%
  \BibitemOpen
  \bibfield  {author} {\bibinfo {author} {\bibfnamefont {F.}~\bibnamefont
  {Michaud}}, \bibinfo {author} {\bibfnamefont {F.}~\bibnamefont {Vernay}},
  \bibinfo {author} {\bibfnamefont {S.~R.}\ \bibnamefont {Manmana}}, \ and\
  \bibinfo {author} {\bibfnamefont {F.}~\bibnamefont {Mila}},\ }\href {\doibase
  10.1103/PhysRevLett.108.127202} {\bibfield  {journal} {\bibinfo  {journal}
  {Phys. Rev. Lett.}\ }\textbf {\bibinfo {volume} {108}},\ \bibinfo {pages}
  {127202} (\bibinfo {year} {2012})}\BibitemShut {NoStop}%
\bibitem [{\citenamefont {Wang}\ \emph {et~al.}(2013)\citenamefont {Wang},
  \citenamefont {Furuya}, \citenamefont {Nakamura},\ and\ \citenamefont
  {Komakura}}]{Wang_3spin}%
  \BibitemOpen
  \bibfield  {author} {\bibinfo {author} {\bibfnamefont {Z.-Y.}\ \bibnamefont
  {Wang}}, \bibinfo {author} {\bibfnamefont {S.~C.}\ \bibnamefont {Furuya}},
  \bibinfo {author} {\bibfnamefont {M.}~\bibnamefont {Nakamura}}, \ and\
  \bibinfo {author} {\bibfnamefont {R.}~\bibnamefont {Komakura}},\ }\href
  {\doibase 10.1103/PhysRevB.88.224419} {\bibfield  {journal} {\bibinfo
  {journal} {Phys. Rev. B}\ }\textbf {\bibinfo {volume} {88}},\ \bibinfo
  {pages} {224419} (\bibinfo {year} {2013})}\BibitemShut {NoStop}%
\bibitem [{\citenamefont {Witten}(1984)}]{Witten_nonabelian_bosonization}%
  \BibitemOpen
  \bibfield  {author} {\bibinfo {author} {\bibfnamefont {E.}~\bibnamefont
  {Witten}},\ }\href {http://dx.doi.org/10.1007/BF01215276} {\bibfield
  {journal} {\bibinfo  {journal} {Commun. Math. Phys}\ }\textbf {\bibinfo
  {volume} {92}},\ \bibinfo {pages} {455} (\bibinfo {year} {1984})}\BibitemShut
  {NoStop}%
\bibitem [{\citenamefont {Di~Francesco}\ \emph {et~al.}(1997)\citenamefont
  {Di~Francesco}, \citenamefont {Mathieu},\ and\ \citenamefont
  {S\'en\'echal}}]{DiFrancesco_CFT}%
  \BibitemOpen
  \bibfield  {author} {\bibinfo {author} {\bibfnamefont {P.}~\bibnamefont
  {Di~Francesco}}, \bibinfo {author} {\bibfnamefont {P.}~\bibnamefont
  {Mathieu}}, \ and\ \bibinfo {author} {\bibfnamefont {D.}~\bibnamefont
  {S\'en\'echal}},\ }\href@noop {} {\emph {\bibinfo {title} {Conformal Field
  Theory}}}\ (\bibinfo  {publisher} {Springer},\ \bibinfo {address} {New
  York},\ \bibinfo {year} {1997})\BibitemShut {NoStop}%
\bibitem [{sup()}]{suppl}%
  \BibitemOpen
  \href@noop {} {}\bibinfo {note} {See Supplemental Material for technical
  details about the modular transformation and the partition
  function.}\BibitemShut {Stop}%
\bibitem [{Note1()}]{Note1}%
  \BibitemOpen
  \bibinfo {note} {In general, the double degeneracy of the ground states in
  the symmetric and antisymmetric sectors with respect to the $g\to -g$ does
  not necessarily mean the spontaneous symmetry breaking: it could correspond
  to a topological degeneracy two ground states in the presence of periodic
  boundary conditions are indistinguishable by any local observable. However,
  in one spatial dimension, which is the focus of the present Letter, such a
  topological order with long-range entanglement is absent~\protect \cite
  {Chen_classification_1D_PRB2011} and thus the ground-state degeneracy between
  the two sectors does imply a spontaneous symmetry breaking.}\BibitemShut
  {Stop}%
\bibitem [{\citenamefont {Zamolodchikov}(1986)}]{Zamolodchikov_c}%
  \BibitemOpen
  \bibfield  {author} {\bibinfo {author} {\bibfnamefont {A.~B.}\ \bibnamefont
  {Zamolodchikov}},\ }\href@noop {} {\bibfield  {journal} {\bibinfo  {journal}
  {JETP Lett.}\ }\textbf {\bibinfo {volume} {43}},\ \bibinfo {pages} {730}
  (\bibinfo {year} {1986})}\BibitemShut {NoStop}%
\bibitem [{\citenamefont {Pradisi}\ \emph
  {et~al.}(1995{\natexlab{a}})\citenamefont {Pradisi}, \citenamefont
  {Sagnotti},\ and\ \citenamefont {Stanev}}]{Pradisi_unori1}%
  \BibitemOpen
  \bibfield  {author} {\bibinfo {author} {\bibfnamefont {G.}~\bibnamefont
  {Pradisi}}, \bibinfo {author} {\bibfnamefont {A.}~\bibnamefont {Sagnotti}}, \
  and\ \bibinfo {author} {\bibfnamefont {Y.}~\bibnamefont {Stanev}},\ }\href
  {http://www.sciencedirect.com/science/article/pii/037026939500532P}
  {\bibfield  {journal} {\bibinfo  {journal} {Phys. Lett.}\ }\textbf {\bibinfo
  {volume} {B354}},\ \bibinfo {pages} {279 } (\bibinfo {year}
  {1995}{\natexlab{a}})}\BibitemShut {NoStop}%
\bibitem [{\citenamefont {Pradisi}\ \emph
  {et~al.}(1995{\natexlab{b}})\citenamefont {Pradisi}, \citenamefont
  {Sagnotti},\ and\ \citenamefont {Stanev}}]{Pradisi_unori2}%
  \BibitemOpen
  \bibfield  {author} {\bibinfo {author} {\bibfnamefont {G.}~\bibnamefont
  {Pradisi}}, \bibinfo {author} {\bibfnamefont {A.}~\bibnamefont {Sagnotti}}, \
  and\ \bibinfo {author} {\bibfnamefont {Y.}~\bibnamefont {Stanev}},\ }\href
  {http://www.sciencedirect.com/science/article/pii/037026939500840H}
  {\bibfield  {journal} {\bibinfo  {journal} {Phys. Lett.}\ }\textbf {\bibinfo
  {volume} {B356}},\ \bibinfo {pages} {230 } (\bibinfo {year}
  {1995}{\natexlab{b}})}\BibitemShut {NoStop}%
\bibitem [{\citenamefont {Ziman}\ and\ \citenamefont
  {Schulz}(1987)}]{ZimanSchulz}%
  \BibitemOpen
  \bibfield  {author} {\bibinfo {author} {\bibfnamefont {T.}~\bibnamefont
  {Ziman}}\ and\ \bibinfo {author} {\bibfnamefont {H.~J.}\ \bibnamefont
  {Schulz}},\ }\href {\doibase 10.1103/PhysRevLett.59.140} {\bibfield
  {journal} {\bibinfo  {journal} {Phys. Rev. Lett.}\ }\textbf {\bibinfo
  {volume} {59}},\ \bibinfo {pages} {140} (\bibinfo {year} {1987})}\BibitemShut
  {NoStop}%
\bibitem [{\citenamefont {Michaud}\ \emph {et~al.}(2013)\citenamefont
  {Michaud}, \citenamefont {Manmana},\ and\ \citenamefont
  {Mila}}]{Michaud_wzw}%
  \BibitemOpen
  \bibfield  {author} {\bibinfo {author} {\bibfnamefont {F.}~\bibnamefont
  {Michaud}}, \bibinfo {author} {\bibfnamefont {S.~R.}\ \bibnamefont
  {Manmana}}, \ and\ \bibinfo {author} {\bibfnamefont {F.}~\bibnamefont
  {Mila}},\ }\href {\doibase 10.1103/PhysRevB.87.140404} {\bibfield  {journal}
  {\bibinfo  {journal} {Phys. Rev. B}\ }\textbf {\bibinfo {volume} {87}},\
  \bibinfo {pages} {140404} (\bibinfo {year} {2013})}\BibitemShut {NoStop}%
\bibitem [{\citenamefont {Simutis}\ \emph {et~al.}(2016)\citenamefont
  {Simutis}, \citenamefont {Gvasaliya}, \citenamefont {Xiao}, \citenamefont
  {Landee},\ and\ \citenamefont {Zheludev}}]{Simutis_Raman_ladder}%
  \BibitemOpen
  \bibfield  {author} {\bibinfo {author} {\bibfnamefont {G.}~\bibnamefont
  {Simutis}}, \bibinfo {author} {\bibfnamefont {S.}~\bibnamefont {Gvasaliya}},
  \bibinfo {author} {\bibfnamefont {F.}~\bibnamefont {Xiao}}, \bibinfo {author}
  {\bibfnamefont {C.~P.}\ \bibnamefont {Landee}}, \ and\ \bibinfo {author}
  {\bibfnamefont {A.}~\bibnamefont {Zheludev}},\ }\href {\doibase
  10.1103/PhysRevB.93.094412} {\bibfield  {journal} {\bibinfo  {journal} {Phys.
  Rev. B}\ }\textbf {\bibinfo {volume} {93}},\ \bibinfo {pages} {094412}
  (\bibinfo {year} {2016})}\BibitemShut {NoStop}%
\bibitem [{\citenamefont {Michaud}\ \emph {et~al.}(2011)\citenamefont
  {Michaud}, \citenamefont {Vernay},\ and\ \citenamefont
  {Mila}}]{Michaud_Raman}%
  \BibitemOpen
  \bibfield  {author} {\bibinfo {author} {\bibfnamefont {F.}~\bibnamefont
  {Michaud}}, \bibinfo {author} {\bibfnamefont {F.}~\bibnamefont {Vernay}}, \
  and\ \bibinfo {author} {\bibfnamefont {F.}~\bibnamefont {Mila}},\ }\href
  {\doibase 10.1103/PhysRevB.84.184424} {\bibfield  {journal} {\bibinfo
  {journal} {Phys. Rev. B}\ }\textbf {\bibinfo {volume} {84}},\ \bibinfo
  {pages} {184424} (\bibinfo {year} {2011})}\BibitemShut {NoStop}%
\bibitem [{\citenamefont {Sato}\ \emph {et~al.}(2012)\citenamefont {Sato},
  \citenamefont {Katsura},\ and\ \citenamefont {Nagaosa}}]{Sato_Raman}%
  \BibitemOpen
  \bibfield  {author} {\bibinfo {author} {\bibfnamefont {M.}~\bibnamefont
  {Sato}}, \bibinfo {author} {\bibfnamefont {H.}~\bibnamefont {Katsura}}, \
  and\ \bibinfo {author} {\bibfnamefont {N.}~\bibnamefont {Nagaosa}},\ }\href
  {\doibase 10.1103/PhysRevLett.108.237401} {\bibfield  {journal} {\bibinfo
  {journal} {Phys. Rev. Lett.}\ }\textbf {\bibinfo {volume} {108}},\ \bibinfo
  {pages} {237401} (\bibinfo {year} {2012})}\BibitemShut {NoStop}%
\bibitem [{\citenamefont {Chen}\ \emph {et~al.}(2011)\citenamefont {Chen},
  \citenamefont {Gu},\ and\ \citenamefont
  {Wen}}]{Chen_classification_1D_PRB2011}%
  \BibitemOpen
  \bibfield  {author} {\bibinfo {author} {\bibfnamefont {X.}~\bibnamefont
  {Chen}}, \bibinfo {author} {\bibfnamefont {Z.-C.}\ \bibnamefont {Gu}}, \ and\
  \bibinfo {author} {\bibfnamefont {X.-G.}\ \bibnamefont {Wen}},\ }\href
  {\doibase 10.1103/PhysRevB.83.035107} {\bibfield  {journal} {\bibinfo
  {journal} {Phys. Rev. B}\ }\textbf {\bibinfo {volume} {83}},\ \bibinfo
  {pages} {035107} (\bibinfo {year} {2011})}\BibitemShut {NoStop}%
\end{thebibliography}

\begin{table*}[t!]
{\bf \large Supplementary Material: \\
``Symmetry Protection of Critical Phases and a Global Anomaly in $1+1$ Dimensions''}
\end{table*}

\newpage

\section{Modular transformations}

Here we briefly explain the physical role of the modular transformations
\begin{align}
 \mathcal{T}: & \quad \tau \to \tau + 1, 
 \label{modular_T} \\
 \mathcal{S}: & \quad \tau \to - \frac 1{\tau}.
 \label{modular_S}
\end{align}
Let us impose the following boundary conditions in two directions of the torus on $g(z,\bar z)$:
\begin{align}
 g(z+1, \bar z+1) &= h_x g(z,\bar z),
 \label{bc_1} \\
 g(z+\tau, \bar z + \bar \tau) &= h_t g(z, \bar z).
 \label{bc_t}
\end{align}
The former corresponds to the boundary condition in the spatial direction and the latter to the boundary condition in the temporal direction.
$h_x$ and $h_t$ are elements of a discrete group, say $H$.
We note that $H$ is a commutative group so that the boundary conditions are consistent.
Note that, when we go around the temporal direction,
we impose the spatial shift by $\operatorname{Re}\tau$ simultaneously.
The periodic boundary condition corresponds to $H=\{1\}$.
The antiperiodic boundary condition in the spatial direction means $h_x = -1$ and $H=\{1,-1\}$.
A torus with another modulus,
\begin{equation}
 \tau' = \frac{a\tau+b}{c\tau+d},
\end{equation}
with $a,b,c,d\in \mathbb Z$ and $ad-bc=1$ is identical that with the modulus $\tau$.
The modular transformation $\tau \to \tau'$ modifies the boundary conditions as
\begin{align}
 g(z+1, \bar z+1) &= h_t^ch_x^d g(z,\bar z),
 \label{bc_1'}\\
 g(z+\tau, \bar z + \bar \tau) &= h_t^a h_x^bg(z,\bar z).
 \label{bc_t'}
\end{align}
Let us see these step by step.
It immediately follows from Eqs.~\eqref{bc_1} and \eqref{bc_t} that
\begin{align}
 g(z+\tau+1, \bar z+\bar \tau +1)
 &= h_th_x g(z,\bar z).
 \label{bc_t+1}
\end{align}
Next, putting together Eqs.~\eqref{bc_1} and \eqref{bc_t+1}, we find that the modular transformation $\tau \to \tau +1$
modifies the temporal boundary condition $h_t$ to $h_th_x$.
Likewise we obtain
\begin{align}
 g(z+a\tau +b, \bar z+a\bar\tau + b)
 & = h_t^a h_x^b g(z,\bar z),
 \label{bc_at+b} \\
 g(z+c\tau+d, \bar z +c\bar\tau +d)
 &= h_t^c h_x^d g(z, \bar z).
 \label{bc_ct+d}
\end{align}
where $a,b,c,d\in \mathbb Z$.
Here we pay attention to a fact that $g(z, \bar z)$ is a primary field.
The primary field is transformed under the conformal transformation $z\to w$ as follows~\cite{DiFrancesco_CFT}.
\begin{equation}
 g(w,\bar w) = \biggl(\frac{dw}{dz}\biggr)^{\Delta}\biggl(\frac{d\bar w}{d\bar z}\biggr)^{\bar \Delta} g(z,\bar z), \quad (\Delta, \bar\Delta>0).
\end{equation}
Considering the conformal transformation $w=(c\tau+d)z$, we are able to rewrite Eqs.~\eqref{bc_at+b} and \eqref{bc_ct+d} in the following form:
\begin{align}
 g\bigl(w+\tfrac{a\tau+b}{c\tau+d}, \bar w+\tfrac{a\bar\tau+b}{c\bar\tau+d}\bigr) &= h_t^ah_x^b g(w,\bar w), \\
 g(w+1,\bar w+1) &= h_t^ch_x^d g(w, \bar w).
\end{align}
They are identical to the boundary conditions \eqref{bc_1'} and \eqref{bc_t'}.

We can rephrase the modular transformations \eqref{modular_T} and \eqref{modular_S} in terms of relations between the factors $h_t$ and $h_x$ as follows.
\begin{align}
 \mathcal{T}:& \quad (h_t, h_x) \to (h_th_x, h_x),
 \label{T} \\
 \mathcal{S}:& \quad (h_t, h_x) \to (h_x^{-1}, h_t).
 \label{S}
\end{align}
Therefore, we may say that the modular transformation $\mathcal{T}$ yields the spatial twist $h_x$ in the temporal boundary condition and that the modular transformation $\mathcal{S}$ exchanges the role of the space and imaginary-time directions.

\section{Kac-Moody characters of WZW$_k$ theory
and their modular transformations}

The Kac-Moody character $\chi_j(\tau)$ of the SU(2)$_k$ WZW theory is
explicitely given by~\cite{Gepner_wzw}
\begin{align}
 \chi_j(\tau) &= \frac{C_{2j+1, k+2}(\tau)}{C_{1,2}(\tau)}, \\
 C_{n,k}(\tau)&= \Theta_{n,k}(\tau) - \Theta_{-n,k}(\tau), \\
 \Theta_{n,k}(\tau)
 &= \sum_{n\in \mathbb Z + \frac j{2k}} e^{i2\pi k n^2\tau}.
\end{align}
The modular transformations of the Kac-Moody characters read
\begin{equation}
\mathcal{T}\chi_j(\tau)=e^{i2\pi(\Delta_j-\frac{c}{24})}\chi_j(\tau) ,
\end{equation}
and 
\begin{equation}
\mathcal{S}\chi_j(\tau)=\sum_{j'}S_{j,j'}\chi_{j'}(\tau), 
\end{equation}
where $c=3k/(2+k)$, $\Delta_j=j(j+1)/(2+k)$, and
\begin{equation}
S_{j,j'}=\sqrt{\frac{2}{k+2}}
\sin\biggl[\frac{\pi(2j+1)(2j'+1)}{k+2}\biggr]. 
\end{equation}
The character of the antiholomorphic part is similarly transformed.

\section{Partition function of the $\mathbb{Z}_2$ orbifold}

Here we present an elementary derivation of
the partition function of the $\mathbb Z_2$ orbifold of WZW$_k$.
To this end the relation (5) in the main text is useful.
The partition function of the original SU(2)$_k$ WZW theory
is simply given in the diagonal form,
\begin{equation}
 Z_{\mathrm{SU(2)}}(\tau,\bar \tau) = \sum_{j=0}^{\frac k2}|\chi_j(\tau)|^2.
  \label{Z_SU2}
\end{equation}
The highest weight state $|j,j\rangle$ of the conformal tower
yielding the term $|\chi_j(\tau)|^2$ in Eq.~\eqref{Z_SU2} 
corresponds to a primary field $g^{2j}$,
which is transformed by $g\to-g$ as follows.
\begin{equation}
 |j,j\rangle \to (-1)^{2j}|j,j\rangle.
\end{equation}
Here we implicitely assumed that the state $|0,0\rangle$ is transformed
as $|0,0\rangle\to|0,0\rangle$ because $|0,0\rangle$ corresponds
to the identity operator and it is even with respect to
the $g\to-g$ transformation.

The conformal tower is composed of the highest weight state and
idescendant states.
The latter are generated from the former
by multiplying generators of the Kac-Moody and the SU(2) current algebras.
Since these generators are odd under $g\to-g$~\cite{DiFrancesco_CFT},
every state generated from $|j,j\rangle$ acquires the nontrivial
phase $(-1)^{2j}$ by the $g\to-g$ transformation.

Now we are ready to derive the projected partition function $Z_+^{\rm proj}$.
The definition (4) in the main text immediately gives
\begin{equation}
 Z_+^{\rm proj} = \sum_{\substack{j=0 \\ j\in\mathbb Z}}^{\lfloor\frac{k}2\rfloor}|\chi_j|^2.
\end{equation}
The index $j$ runs over integers only and $\lfloor \frac k2 \rfloor$ denotes a maximum integer that satisfies $\lfloor \frac k2 \rfloor \le \frac k2$.
We apply $\mathcal S$ and $\mathcal{TS}$ transformations on $Z_+^{\rm proj}$:
\begin{align}
 \mathcal SZ_+^{\rm proj}(\tau)
 &= \sum_{\substack{j=0 \\ j\in\mathbb Z}}^{\lfloor\frac{k}2\rfloor}|S_{j,0}\chi_0 + S_{j,\frac 12}\chi_{\frac 12} + \cdots + S_{j,\frac k2}\chi_{\frac k2}|^2, \\
 \mathcal T\mathcal SZ_+^{\rm proj}(\tau)
 &= \sum_{\substack{j=0 \\ j\in\mathbb Z}}^{\lfloor\frac{k}2\rfloor} |S_{j,0}\chi_0 + S_{j,\frac 12}e^{i2\pi\Delta_{1/2}}\chi_{\frac 12} +  \cdots
 \notag \\
 & \qquad
 + S_{j,\frac k2}e^{i2\pi\Delta_{k/2}}\chi_{\frac k2}|^2.
\end{align}
Combining them with the formula (5) in the main text, 
we obtain the coefficient $X_{j,j'}$ of Eq.~(2) in the main text.
For $j=j'$ and $j=0,1,2,\cdots, \lfloor \frac{k}2\rfloor$,
\begin{equation}
 X_{j,j} = 2(S_{0,j}^2+S_{1,j}^2+\cdots + S_{\lfloor \frac{k}2 \rfloor,j}^2).
  \label{X_diag_int}
\end{equation}
For $j=j'$ and $j=\frac 12, \frac 32, \cdots, \lfloor\frac{k-1}2\rfloor+\frac 12$,
\begin{equation}
 X_{j,j} = 2(S_{0,j}^2+S_{1,j}^2+\cdots + S_{\lfloor\frac{k}2\rfloor,j}^2)-1.
  \label{X_diag_half_int}
\end{equation}
For off-diagonal terms,
\begin{align}
 X_{j,j'} 
 &= (S_{0,j}S_{0,j'}+S_{1,j}S_{1,j'}+\cdots + S_{\lfloor\frac{k}2\rfloor,j}S_{\lfloor\frac{k}2\rfloor, j'})
 \notag \\
 & \qquad \times
 (1+e^{i2\pi(\Delta_j-\Delta_{j'})}).
 \label{X_off_diag}
\end{align}
The diagonal coefficients \eqref{X_diag_int} and \eqref{X_diag_half_int} are simple.
The summation $2(S_{0,j}^2+S_{1,j}^2+\cdots +S_{\lfloor\frac{k}2\rfloor,j}^2)$ becomes
\begin{align}
 &2(S_{0,j}^2+S_{1,j}^2+\cdots +S_{\lfloor\frac{k}2\rfloor,j}^2) \notag \\
 &= \frac 4{2+k}\sum_{n=0}^{\lfloor\frac{k}2\rfloor} \sin^2\biggl(\frac{\pi(2j+1)(2n+1)}{2+k}\biggr) \notag \\
 &= \frac 2{2+k} \biggl[ \biggl\lfloor \frac k2 \biggr\rfloor + 1 
 - \sum_{n=0}^{\lfloor\frac{k}2\rfloor}\cos\biggl(\frac{2\pi(2j+1)(2n+1)}{2+k}\biggr)\biggr] \notag \\
 &= \frac 2{2+k}\biggl( \biggl\lfloor \frac k2 \biggr\rfloor + 1 -\frac{\sin[\theta(2\lfloor \frac k2 \rfloor+2)]}{2\sin \theta}\biggr),
\end{align}
with $\theta = 2\pi(2j+1)/(2+k)$.
The identity $\sin[\theta(k+1)]=\sin(2\pi(2j+1)-\theta)=-\sin\theta$ 
leads to
\begin{equation}
 X_{j,j}=1, \quad (j=0,1,2,\cdots, \lfloor\tfrac{k}2\rfloor)
\end{equation}
and
\begin{equation}
 X_{j,j}=0, \quad (j=\tfrac 12, \tfrac 32, \cdots, \lfloor\tfrac{k-1}2\rfloor + \tfrac 12).
\end{equation}

The off-diagonal term is similarly derived.
\begin{align}
 &S_{0,j}S_{0,j'}+S_{1,j}S_{1,j'} + \cdots + S_{\frac{k-1}2,j}S_{\lfloor\frac{k}2\rfloor,j'}  \notag \\
 &=\frac 1{2+k} \sum_{n=0}^{\lfloor\frac{k}2\rfloor} \biggl[ - \cos\biggl(\frac{\pi(2j+2j'+2)(2n+1)}{2+k}\biggr)
 \notag \\
 & \qquad + \cos \biggl(\frac{\pi(2j-2j')(2n+1)}{2+k}\biggr)\biggr].
\end{align}
When $j+j'\not=\frac k2$, the summation in the last line leads to
\begin{align}
 &S_{0,j}S_{0,j'}+S_{1,j}S_{1,j'} + \cdots + S_{\lfloor\frac{k}2\rfloor,j}S_{\lfloor\frac{k}2\rfloor,j'} \notag \\
 &= \frac 1{2+k}\biggl[ -\frac{\sin[\theta_+(2\lfloor \frac k2\rfloor +2)]}{2\sin\theta_+} + \frac{\sin[\theta_-(2\lfloor \frac k2\rfloor +2)]}{2\sin\theta_-}
 \biggr] \notag \\
 &= 0,
\end{align}
where $\theta_+=\pi(2j+2j'+2)/(2+k)$ and $\theta_-=\pi(2j-2j')/(2+k)$.
On the other hand, when $j+j'=\frac k2$, the summation does not vanish.
\begin{align}
 &S_{0,j}S_{0,j'}+S_{1,j}S_{1,j'} + \cdots + S_{\lfloor\frac{k}2\rfloor,j}S_{\lfloor\frac{k}2\rfloor,j'} \notag \\
 &= \frac 1{2+k} \biggl(\biggl\lfloor \frac k2 \biggr\rfloor + 1 +\frac{\sin[\theta_-(2\lfloor \frac k2 \rfloor + 2)]}{2\sin \theta_-} \biggr) \notag \\
 &= \frac 12.
\end{align}
In the end, the nonzero coefficients $X_{j,j'}$ are
\begin{equation}
 X_{j,j'}= \left\{
  \begin{array}{ccl}
   1, & & (j=0,1, \cdots, \lfloor\frac{k}2\rfloor), \\
   \frac 12(1+e^{i2\pi(\Delta_j-\Delta_{j'})}), & & (j+j'=\frac k2).
  \end{array}\right.
 \label{X}
\end{equation}
$X_{j,\frac k2 - j}$ is not real for odd $k$  because
\begin{align}
 e^{i2\pi(\Delta_j-\Delta_{\frac k2-j})}
 &= e^{i2\pi(j-\frac k4)} 
 = (-1)^{2j}(-i)^k.
\end{align}

\subsection{$k\equiv 0$ mod $4$}

When $k$ is divisible by $4$, the coefficient \eqref{X} becomes
\begin{equation}
 X_{j,j'}= \left\{
  \begin{array}{ccl}
   1, & & (j=j' \text{ and } j=0,1, \cdots, \frac{k}2), \\
   \frac 12[1+(-1)^{2j}], & & (j+j'=\frac k2), \\
   0, & & \text{otherwise},
  \end{array}\right.
 \label{X_4n}
\end{equation}
resulting in the partition function,
\begin{equation}
 Z_+ = \sum_{\substack{j=0 \\ j\in\mathbb Z}}^{\frac{k-4}4} |\chi_j+\chi_{\frac k2-j}|^2 + 2|\chi_{\frac k4}|^2.
  \label{Z_+_4n}
\end{equation}
Note that the term $|\chi_{\frac k4}|^2$ comes from the first
the second lines of Eq.~\eqref{X_4n}.

\subsection{$k\equiv 2$ mod $4$}

When $k$ is even but not divisible by $4$, Eq.~\eqref{X} becomes
\begin{equation}
 X_{j,j'}= \left\{
  \begin{array}{ccl}
   1, & & (j=j' \text{ and } j=0,1, \cdots, \frac{k-1}2), \\
   \frac 12[1-(-1)^{2j}], & & (j+j'=\frac k2), \\
   0, & & \text{otherwise}.
  \end{array}\right.
 \label{X_4n-2}
\end{equation}
That is, the partition function is given by
\begin{align}
 Z_+
 &= \sum_{\substack{j=0 \\ j\in\mathbb Z}}^{\frac k2}|\chi_j|^2 
 + |\chi_{\frac k4}|^2
  + \sum_{\substack{j=1/2 \\ j\in\mathbb Z+1/2}}^{\frac{k-2}4} (\chi_j \bar \chi_{\frac k2-j}+\chi_{\frac k2-j}\bar\chi_j).
  \label{Z_+_4n-2}
\end{align}

\subsection{$k\equiv 1$ mod $2$}

When $k$ is odd, the coefficient is not real:
\begin{equation}
 X_{j,j'}= \left\{
  \begin{array}{ccl}
   1, & & (j=0,1, \cdots, \frac{k-1}2), \\
   \frac 12[1+(-1)^{2j}e^{-i\pi k/2}], & & (j+j'=\frac k2), \\
   0, & & \text{otherwise}.
  \end{array}\right.
 \label{X_2n-1}
\end{equation}
The partition function is then anomalous,
\begin{align}
 Z_+
 &= \sum_{\substack{j=0 \\ j\in\mathbb Z}}^{\frac k2}|\chi_j|^2  
  + \sum_{j=0}^{\frac k2} \biggl(\frac{1+(-1)^{2j}e^{-i\pi k/2}}2\chi_j \bar \chi_{\frac k2-j}
 \notag \\
 & \quad
 +\frac{1+(-1)^{2j}e^{i\pi k/2}}2\chi_{\frac k2-j}\bar\chi_j\biggr).
  \label{Z_+_2n-1}
\end{align}

\subsection{Modular invariance}

The modular noninvariance of the odd-level $\mathbb Z_2$ orbifold \eqref{Z_+_2n-1} is obvious because, for instance, $\chi_j\bar\chi_{\frac k2-j}$
is transformed as
\begin{equation}
 \mathcal T(\chi_j\bar\chi_{\frac k2-j})
  = (-1)^{2j}e^{-i\pi k/4}\chi_j\bar\chi_{\frac k2-j}.
\end{equation}
It is also straightforward to confirm the modular invariance
of the even-level $\mathbb Z_2$ orbifolds \eqref{Z_+_4n} and \eqref{Z_+_4n-2}.
According to a complete classification of all the possible modular invariants in the SU(2) WZW theory~\cite{DiFrancesco_CFT},
the partition function \eqref{Z_+_4n} for $k=4n$ is the modular invariant
of the type $D_{2n+2}$ and one \eqref{Z_+_4n-2} for $k=4n-2$
is that of the type $D_{2n+1}$.

\section{Symmetry protection by parity (space inversion) symmetry}

The $g\to-g$ symmetry is not the only symmetry that protects the
two categories of the SU(2)- and Lorentz-invariant critical behaviors.
The parity symmetry
$g(z,\bar{z})\to-g^{-1}(\bar{z},z)$ also protects them because the CFT
obtained by gauging the $g(z,\bar{z})\to-g^{-1}(\bar{z},z)$ symmetry has
the same spectrum with the untwisted sector of the $\mathbb{Z}_2$ orbifold of the $g\to-g$ symmetry~\cite{Pradisi_unori1,Pradisi_unori2}.  
There is another parity symmetry $g(z,\bar{z})\to g^{-1}(\bar{z},z)$.
However, this does not protect the two categories unlike
the previous one.
In the spin chain context,
the $g(z,\bar{z})\to-g^{-1}(\bar{z},z)$ parity, which
is ``protecting'', corresponds to the
site-centered inversion $\bm{S}_j\to\bm{S}_{-j}$.
On the other hand, either the time reversal
$\bm{S}_j\to-\bm{S}_j$ or the bond-centered inversion
$\bm{S}_{j}\to\bm{S}_{1-j}$ symmetry yields
the ``unprotecting'' parity symmetry,
$g(z,\bar{z})\to g^{-1}(\bar{z},z)$.

\section{Correspondence between the WZW$_k$ and its $\mathbb{Z}_2$ orbifold}

Here we show that there exists a correspondence between the WZW$_k$ and its $\mathbb{Z}_2$ orbifold in a sense that 
a $\mathbb{Z}_2$ orbifold of the $\mathbb{Z}_2$ orbifold is identical to WZW$_k$.
Let us denote the $\mathbb{Z}_2$ orbifold of WZW$_k$ as O$_k$.
The correspondence between WZW$_k$ and O$_k$ can also be seen as a correspondence of renormalization group (RG) flows.
Namely, if there is an RG flow between WZW$_k$ and WZW$_{k'}$, a corresponding RG flow exists between 
O$_k$ and O$_{k'}$ and vice versa.

To show the correspondence, we recall a fact that the partition function $Z_+$ of O$_k$ [Eq.~(5) in the main text] is given by
\begin{equation}
 Z_+ = \frac 12 \sum_{h_x, h_t\in \{1,-1\}} Z_{(h_x,h_t)},
  \label{Z_+}
\end{equation}
where $Z_{(h_x,h_t)}$ is the partition function of WZW$_k$ under boundary conditions,
\begin{equation}
 \left\{
 \begin{split}
  g(z+1,\bar z+1) &= h_x g(z,\bar z), \\
  g(z+\tau, \bar z + \bar \tau) &= h_t g(z, \bar z).
 \end{split}
 \right.
\end{equation}
The projected partition function $Z_+^{\mathrm{proj}}$ [Eq.~(4) in the main text] is represented as
\begin{equation}
 Z_+^{\mathrm{proj}} = \frac 12 [Z_{(1,1)} + Z_{(1,-1)}].
\end{equation}
Recalling the operations \eqref{T} and \eqref{S} of the modular transformations, we can show that Eq.~\eqref{Z_+} equals to Eq.~(5) in the main text.

Now we take a $\mathbb{Z}_2$ orbifold of O$_k$, considering the following $\mathbb{Z}_2$ operation.
First we need to recall a fact that the spatially twisted sector of the O$_k$, which is $[Z_{(-1,1)}+Z_{(-1,-1)}]/2$ in $Z_+$, contains
a soliton field $\psi$~\cite{Gepner_wzw}.
In other words, insertion of the field $\psi$ to the torus transforms the spatial boundary condition from $h_x$ to $-h_x$.
By definition, the $\psi$ field has a $\mathbb Z_2$ nature: the product of $\psi$ with itself gives identity.
Considering an inversion operation $\mathcal{I}:\psi \to - \psi$, 
we can take a $\mathbb{Z}_2$ orbifold of O$_k$ with respect to the $\mathbb{Z}_2$ symmetry $\mathcal{I}$.
As we did in defining O$_k$, we can consider $Z'_{(h_x,h_t)}$.
In particular, $Z'_{(1,1)}=Z_+$.
The partition function $Z'_+$ of the $\mathrm{Z}_2$ orbifold of O$_k$ is given by
\begin{equation}
 Z'_+ = \frac 12 \sum_{h_x,h_t\in\{1,-1\}} Z'_{(h_x,h_t)}.
  \label{Z'_+}
\end{equation}
To investigate $Z'_+$, we focus on the following relations that follow from the definition of $\mathcal{I}$:
\begin{equation}
 \left\{
  \begin{split}
   \mathcal{I} [Z_{(1,1)}+Z_{(1,-1)}] &= Z_{(1,1)} + Z_{(1,-1)}, \\
   \mathcal{I} [Z_{(-1,1)} + Z_{(-1,-1)}] &= - [Z_{(-1,1)} + Z_{(-1,-1)}].
  \end{split}
 \right.
\end{equation}
$Z'_{(1,-1)}$, $Z'_{(-1,1)}$ and $Z'_{(-1,-1)}$ are derived from $Z'_{(1,1)}$ as follows.
\begin{align}
 Z'_{(1,-1)}
 &= \mathcal{I} Z'_{(1,1)}
 \notag \\
 &= \frac 12 \bigl[ Z_{(1,1)} + Z_{(1,-1)} - Z_{(-1,1)} - Z_{(-1,-1)}\bigr],
 \label{Z'_{(1,-1)}}\\
 Z'_{(-1,1)}
 &= \mathcal{S} Z'_{(1,-1)}
 \notag \\
 &= \frac 12 \bigl[ Z_{(1,1)} + Z_{(-1,1)} - Z_{(1,-1)} - Z_{(-1,-1)}\bigr],
 \label{Z'_{(-1,1)}}\\
 Z'_{(-1,-1)}
 &= \mathcal{T}Z'_{(-1,1)}
 \notag \\
 &= \frac 12 \bigl[ Z_{(1,1)} + Z_{(-1,-1)} - Z_{(1,-1)} - Z_{(-1,1)}\bigr].
 \label{Z'_{(-1,-1)}}
\end{align}
Combining Eqs.~\eqref{Z'_{(1,-1)}}, \eqref{Z'_{(-1,1)}} and \eqref{Z'_{(-1,-1)}} with $Z'_{(1,1)}=Z_+$, we obtain
\begin{equation}
 Z'_+ = Z_{(1,1)}.
\end{equation}
Therefore, the $\mathbb{Z}_2$ orbifold of O$_k$ is WZW$_k$.

\end{document}